\documentclass[12pt]{article}

\usepackage{amsmath}
\usepackage{amssymb}
\usepackage{amsfonts}

\def\d{{\rm d}}
\def\im{{\rm i}}
\def\pd#1#2{\frac{\partial#1}{\partial#2}} 

\def\H{{\cal H}}
\def\R{{\cal R}}
\def\HQ{{\cal H}_\textrm{Q}}
\def\HD{{\cal H}^\textrm{D}}
\def\HK{{\cal H}^\textrm{K}}
\def\HQD{{\cal H}_\textrm{Q}^\textrm{D}}
\def\HQK{{\cal H}_\textrm{Q}^\textrm{K}}
\def\SQ{S_\textrm{Q}}
\def\SQD{S_\textrm{Q}^\textrm{D}}
\def\SQK{S_\textrm{Q}^\textrm{K}}

\def\ave#1{\left\langle#1\right\rangle}
\def\av#1{\langle#1\rangle}

\def\ang#1#2{\langle#1,#2\rangle}

\def\abs#1{\left|#1\right|}
\def\T{{T}}
\def\P{{\Pi}}
\def\sch{Schr\"odinger}
\def\kuchar{Kucha\v{r}}
\def\haj{H\'aj\'i\v{c}ek}
\def\schon{Sch\"on}

\begin{document}

\title{An action principle for the quantization of parametric theories and nonlinear quantum cosmology}


\author{Charles H-T Wang, Smaragda Kessari and Edward R Irvine\\
Department of Physics, Lancaster University\\
Lancaster LA1 4YB, UK\\
E-mail: c.wang@lancaster.ac.uk,\\ s.kessari@lancaster.ac.uk, e.irvine@lancaster.ac.uk}

\date{}

\maketitle

\abstract{By parametrizing the action integral for the standard
\sch{} equation  we present a derivation of the recently proposed
method for quantizing a parametrized theory.  The reformulation
suggests a natural extension from conventional to nonlinear
quantum mechanics. This generalization enables
a unitary description of
the quantum
evolution for a broad class of constrained Hamiltonian systems with
a nonlinear kinematic structure. In particular, the new theory is
applicable to the quantization of cosmological models where a
chosen gravitational degree of freedom acts as geometric time.
This is demonstrated explicitly using three cosmological models:
the Friedmann universe with a massless scalar field and Bianchi
type I and IX models. Based on these investigations, the prospect of
further developing the proposed quantization scheme in the context
of quantum gravity is discussed.}

\vskip 7mm


\vskip 10mm

\section{Introduction}

The Hamiltonian formalism and canonical quantization of Einstein's
theory of gravitation was originally developed by Dirac
\cite{Dirac}, Arnowitt, Deser, Misner (ADM) \cite{ADM_1962},
Wheeler \cite{Wheeler}, DeWitt \cite{DeWitt_1967} and others. The
Dirac-ADM action for general relativity is constructed by
representing the spacetime metric using the  spatial metric
components together with the lapse and shift functions that
specify a time foliation. The latter set of functions enter into
the action principle as Lagrangian multipliers ensuring the
general covariance of the theory. Indeed, the `already
parametrized' form of the Dirac-ADM action requires that not all
three-metric components and their conjugate momenta are to be
freely specified  on a spatial Cauchy surface. The Hamiltonian and
momentum constraints must be satisfied initially and will be
dynamically preserved \cite{TEITELBOIM}. A major challenge in
quantizing general relativity is the treatment of these
constraints. The straightforward application of the Dirac
constraint quantization leads to a functional form of the
Klein-Gordon equation, the Wheeler-DeWitt equation, and the
associated difficulties in admitting a probabilistic
interpretation. An alternative approach, the ADM quantization
method, attempts to circumvent these problems by isolating four
out of the six spatial metric components (and their momenta) and
regarding them as the evolution and coordinatization parameters,
known as the {\em embedding variables} \cite{Wheeler_1962,
kuchar_1972, kuchar_1976}, for the quantum evolution of the
remaining `true' gravitational degrees of freedom. However
different set of embedding variables can result in inequivalent
quantizations. A further drawback of this approach is that the
reduced Hamiltonian operator may involve a square root of an
indefinite expression and therefore may not be Hermitian.

The above pioneering work was followed by \kuchar's  significant
contributions to canonical quantum gravity in which  Dirac
quantization and the use of embedding variables are assimilated
\cite{kuchar_1971, kuchar_1972, kuchar_1976}. Guided by the
analogy with parametrized field theory, a canonical transformation
is sought such that in the new set of variables the Hamiltonian
and momentum constraints are linear in the conjugate momenta of
the embedding variables \cite{kuchar_1981, kuchar_1992}.
Remarkably, if this `\kuchar{} transformation' exists then Dirac
and ADM quantization in fact coincide, with both methods leading
to the same functional \sch{} equation for the quantum evolution
of the non-embedding variables. Furthermore, this approach {holds
out} the possibility of constructing a framework for quantum
geometrodynamics with the genuine Lie algebra of spacetime
diffeomorphism. This was meticulously investigated by Isham and
\kuchar{} \cite{isham_kuchar_1985}. A stringent criterion for
\kuchar's embedding variables is that they must be spacetime
scalars, to guarantee the integrability of quantum evolution and
its independence of time slicing \cite{kuchar_1992}. Nevertheless,
\kuchar{} successfully demonstrated that in certain
midisuperspace models, i.e. simple {\em field-theoretical} models
with Killing symmetries, it is possible to find embedding
variables explicitly that fulfill this requirement
\cite{kuchar_1971, kuchar_1994}. Unfortunately, for some other
important models, notably the `scalar geons' \cite{Berger_1972,
Lund_1973, Unruh_1976, Romano}, finding the embedding variables has
proven to be a formidable task.


Recently, \haj{} and Kijowski developed a new approach in which
the local existence of covariant gauge fixings is exploited to
effect the decomposition of canonical variables {\em on the
constraint surface} (defined by the Hamiltonian and momentum
constraints) into embedding and dynamical variables
\cite{hajicek_1999, HK}. The resulting `\kuchar{} decomposition'
may form a basis for a complete \kuchar{} transformation for a
large class of geometries. The usefulness of this approach has
been demonstrated by \haj{} and Kiefer in studying the quantum
dynamics of a collapsing shell of null dust with spherical
symmetry \cite{dust} derived from the `Louko-Whiting-Friedman
model' \cite{LWF}. However if a full theory of quantum gravity is
to be constructed based on the covariant gauge fixing description
then some obstructions will be encountered as have been noted in
\cite{HK}: A complete \kuchar{} transformation becomes
unattainable for geometries with Killing symmetries, which has
also been observed by Torre \cite{torre_1992}. Besides, different
gauges can lead to inequivalent quantum theories \cite{torre_1999}
and global gauge fixing may not exist at all \cite{hajicek_1986,
hajicek_1990}.

The purpose of this paper is to present an alternative approach to
the quantization of a parametric theory, with a view to quantizing
gravity, that will complement the above developments. While much
insight has been gained from the quantization of a theory
parametrized at the classical level, guidance for quantum gravity
gleaned this way appears to be limited. Apart from technical
difficulties and topological restrictions of the explicit
constructions of \kuchar's embedding variables, there is a
conceptual issue concerning these {\em preferred} variables. This
is part of what has been attributed to the `problem(s) of time'
\cite{kuchar_1992, isham_1993}.
More specifically, the existence
of a functional time relies on the canonical transformation for
the geometrodynamical variables:
$ (g_{ab}, p^{ab}) \rightarrow ({X}^A,P_A,q^r,p_r)$ that
splits the Hamiltonian and momentum constraints
$\H_A({X}^B,P_B,q^r,p_r)$ into a `kinematic part'
$\HK_A({X}^B,P_B)$ and a `dynamic part' $\HD_A({X}^B,q^r,p_r)$,
i.e.
\begin{equation}\label{}
\H_A({X}^B,P_B,q^r,p_r) =  \HK_A({X}^B,P_B) + \HD_A({X}^B,q^r,p_r) .
\end{equation}
(Here $a,b=1,2,3;\;A,B=0,1,2,3;\;r=1,2$.)
A crucial assumption made by  \kuchar{} is that $\HK_A({X}^B,P_B)$ are {\em linear}
in $P_A$ so that the theory is `deparametrizable' with ${X}^A$ and
$q^r$ identified as embedding and dynamical variables
respectively. However an obstacle typical of midisuperspace models
that prevents deparametrization is the nonlinear nature of
$\HK_A({X}^B,P_B)$. Although in some cases,
an ingenious choice
of $X^A$ may well be the way out, this can be technically
infeasible and might not be possible in principle. In these
circumstances, we adopt the viewpoint that it may be possible to
formulate a parametric quantum evolution of the dynamical
variables $q^r$  and treat the remaining variables $X^A$ as
unquantized kinematic variables without further restrictions. The
new formulation sought should, nevertheless, reduce to Dirac
quantization and is expected to be equivalent to
certain other
schemes  up to a factor ordering \cite{bar}
{(in a semiclassical limit)}
if  \kuchar{}
transformation exists that reduces
$\HK_B({X}^A,P_A)$ to simply $P_A$.

As a first step towards the fully quantum field-theoretical
formulation of our approach, a quantum mechanical description is
adopted. In this simplified picture it is possible to focus on
conceptual issues while establishing the essential methodology
with a view to further generalization. Following a recent
publication \cite{wang_2003} where a new quantization scheme was
proposed, we report in this paper how the proposed scheme can
arise from a direct parametric description of quantum evolution as
opposed to seeking to quantize a parametrized classical theory in
a conventional manner. In section~\ref{parasch}, a parametric form
of quantum evolution equations is introduced based on a
parametrized action principle for the \sch{} equation describing a
non-relativistic particle. The advantage of this reformulation is
that it opens up new avenues for generalized parametric quantum
evolution  beyond the description of a Newtonian particle. The
generalized method is applied in section~\ref{frw} to a simple
Friedmann universe filled with a massless scalar field. Two
further cosmological models, of Bianchi types I and IX, both
involving gravity only are analyzed in sections~\ref{Bianchi}
using the nonlinear quantization scheme. Conclusions and
discussions are made in section \ref{fin}  where speculations on
future work are discussed. Units in which $c = \hbar = 16 \pi G =
1$ are adopted throughout.

\section{Parametrized quantum evolution of a non-relativistic particle}
\label{parasch}

Consider a non-relativistic quantum particle of unit mass moving
in an $n$-dimensional Riemannian manifold $M$ with a
time-dependent metric and coordinates $q^a,\;(1  \le a \le n)$
subject to a potential $V(q^a,T)$. In terms of the Newtonian time
$T$, the weight $\frac12$ wavefunction $\Psi(q^a,T)$  of the
particle satisfies the \sch{} equation
\begin{equation}
\im \pd{{\Psi}}{\T} = \hat{h} \,\Psi
\label{scheq}
\end{equation}
where the $\hat{h}=\hat{h}(T)$ denotes the time-dependent
Hamiltonian operator
\begin{equation}\label{}
\hat{h}(T) = -\frac12\Delta + V
\end{equation}
with $\Delta$ denoting the Laplace-Beltrami operator on weight $\frac12$ functions
on $M$ \cite{kuchar_1981, wang_2003}.

It has long been known that, in terms of the equal time inner
product
\begin{equation}
\label{angPsi}
\ang{\Psi_1}{\Psi_2} := \int  \!\Psi_1^* \Psi_2 \,\d^n\! q
\end{equation}
of wavefunctions $\Psi_1$ and $\Psi_2$, the  \sch{} equation
\eqref{scheq}
can be derived from the action integral (e.g.
\cite{dewitt_1964}):
\begin{equation}\label{schact}
\SQ[\Psi, \Psi^*] =
\int \! \d \T\, \Re \ang{\Psi}{ (\im \partial_{\T} - \hat{h}) \,\Psi}
\end{equation}
by freely varying $\Psi$ and its complex conjugate $\Psi^*$.\footnote{The subscript `Q' indicates `quantum'.}
However the potential significance of this
formulation in the context of constrained Hamiltonian systems does
not appear to have been fully recognized in the literature and
will be explored in this paper. In what follows
an action integral equivalent to \eqref{schact} will be
constructed that generates quantum evolution equations
with respect to a general time coordinate, by parametrizing the
Newtonian time and turning it into a constrained classical
variable.

We anticipate the parametrized $T$ and its momentum to couple with
a reformulated \sch{} equation in a manner similar to the
interaction between classical and quantum variables.     Of
course, the `classical variables' arising from this procedure do
{\em not} carry additional physical degrees of freedom due to the
constraining relations accompanying the parametrized theory.
Therefore these `constrained classical variables' are really
kinematic (embedding) variables, rather than genuine dynamical
variables, forging a link between the physical and geometrical
descriptions of the systems  \cite{wang_2003}. Nonetheless the
expected mathematical similarity with the semiclassical theory
suggests that the action integral of the parametrized quantum
theory is a sum of its kinematic and dynamic parts. While the
kinematic part depends on `classical variables' only the dynamic
part will involve both classical and quantum variables derived
from the original action \eqref{schact}, whose solutions have the
scaling invariance under $\Psi \rightarrow \lambda \Psi$ for any
complex constant $\lambda$, even though $\abs{\lambda} \neq 1$
\cite{kibble_1980}. However this scaling invariance will
apparently be  violated due to the presence of the kinematic part
of the action that is independent of the quantum variables. A way
to avoid this problem is to eliminate the scaling invariance by
using, instead of \eqref{schact}, the modified action integral
\cite{kibble_1980, kibble_1981}:
\begin{equation}\label{schact1}
\SQ[\Psi, \Psi^*, \alpha] =
\int\!\d \T\left\{ \Re \ang{\Psi}{ \im \partial_{\T}  \Psi} -
\ang{\Psi}{ \hat{h} \,\Psi} + \alpha\left(\ang{\Psi}{\Psi} -1    \right)\right\}
\end{equation}
with $\alpha=\alpha(T)$.
With respect to this  Lagrangian multiplier,
the variation
of \eqref{schact1} enforces the normalization
condition
\begin{equation}\label{ncnd}
\ang{\Psi}{\Psi} = 1
\end{equation}
for the wave function $\Psi$, whereas the variation with respect
to $\Psi$ and $\Psi^*$ yields the \sch{} equation of the form
\begin{equation}
\im \pd{{\Psi}}{\T} = (\hat{h}  - \alpha) \,\Psi.
\label{scheq1}
\end{equation}
Despite that  $\alpha$ can be specified arbitrarily, different
choices of it correspond to wavefunctions unitarily related by a
phase depending only on time and   therefore   considered as
physically equivalent.

The action \eqref{schact1} will now be brought to a parametric
form. This
procedure
starts by expressing $T$ as an arbitrarily
chosen function of a parameter time $t$ to yield
\begin{equation}\label{scheq2}
\SQ[\Psi, \Psi^*, \alpha] =
\int\!\d t \left\{ \Re \ang{\Psi}{ \im \partial_{t}  \Psi} -
\dot\T \left[\ang{\Psi}{ \hat{h} \,\Psi} - \alpha\left(\ang{\Psi}{\Psi} -1    \right)\right]\right\}
\end{equation}
where the over dot denotes $\partial_{t}$. Note that in this
expression $\Psi = \Psi(q^a,t) := \Psi(q^a,T(t))$, $\alpha =
\alpha(t) := \alpha(T(t))$ and $\hat{h} = \hat{h}(t) :=
\hat{h}(T(t))$  (with slight abuse of notation). At this stage
$T(t)$ is merely an externally supplied function. It can, however,
be included as part of an enlarged set of dependent variables
augmented with a new pair of Lagrangian multiplier and constraint.
To this end introduce the variable $\P(t)$ and use it to define
the quantity
\begin{equation}\label{}
\HQ := \P + \ang{\Psi}{ \hat{h} \,\Psi} -
\alpha\left(\ang{\Psi}{\Psi} -1    \right)
\end{equation}
so that, if $\HQ$ vanishes, $\P$ equals the coefficient of $\dot{T}$
in the action thereby serving as the momentum of
$T(t)$. By replacing the coefficient of $\dot{T}$ with $\P$ in the
Lagrangian for \eqref{scheq2} and adjoining this Lagrangian with
the term $-N \HQ$ using a Lagrangian multiplier $N=N(t)$, we
are led to the following action
\begin{equation}\label{scheq3}
\SQ[\Psi, \Psi^*, \alpha, \T, \P, N] = \int\!\d t \left\{ \Re
\ang{\Psi}{ \im \partial_{t}  \Psi} + \P\,\dot\T  -N \HQ \right\} .
\end{equation}
This action takes a parametric form and is equivalent to
\eqref{scheq2} upon variation with respect to all components in
the extended set of variables $\{\Psi, \Psi^*, \alpha, \T, \P,N\}$.
It is not hard to further express this action in a complete
canonical form by defining the conjugate momenta of $\Psi$ and
$\Psi^*$. This form  is however not required in our present
analysis.

It is instructive to split $\HQ$  into its `kinematic' part
\begin{equation}\label{HKP}
\HQK = \P
\end{equation}
and `dynamic' part
\begin{equation}\label{}
\HQD = \ang{\Psi}{ \hat{h} \,\Psi} - \alpha\left(\ang{\Psi}{\Psi}
-1    \right)
\end{equation}
so that
\begin{equation}\label{}
\HQ= \HQK + \HQD .
\end{equation}
Accordingly, the  structure of parametrized action \eqref{scheq3}
can be made more transparent with a similar split:
\begin{equation}\label{}
\SQ =
\SQK+
\SQD
\end{equation}
in terms of the kinematic part
\begin{equation}\label{}
\SQK[ \T, \P, N] = \int\!\d t \left\{\P\, \dot\T  -N \HQK \right\}
\end{equation}
and the dynamic part
\begin{equation}\label{}
\SQD[\Psi, \Psi^*, \alpha, \T, \P, N] = \int\!\d t \left\{ \Re
\ang{\Psi}{ \im \partial_{t}  \Psi} -N \HQD \right\} .
\end{equation}
The  parametrized quantum evolution equations can be explicitly
derived from the action \eqref{scheq3} under variations with
respect to $\Psi$ and its conjugate, $\alpha$, $\T$, $\P$ and $N$
to be
\begin{equation}\label{paramscheq}
  \im\partial_t \Psi = N (\hat{h} - \alpha)\,\Psi
\end{equation}
\begin{equation}\label{}
  \ang{\Psi}{\Psi} =1
\end{equation}
\begin{eqnarray}
  \dot\T &=& N \pd{\HQ}{\P} \\[4pt]
  \dot\P &=& -N \pd{\HQ}{\T}
\end{eqnarray}
\begin{equation}\label{paramH}
  \HQ = 0
\end{equation}
respectively. The structure of these equations indeed resembles
that of a semiclassical theory. It is evident that the kinematic
action $\SQK$ involves only `classical' variables and takes a
canonical form. This makes it possible to perform a canonical
transformation, say, from $(\T, \P)$ to  $(\T', \P')$ by
preserving the form of $\SQK[ \T, \P, N] = \SQK[ \T', \P', N]$. It
follows immediately that with these transformed variables, the
action $\SQ[\Psi, \Psi^*, \alpha, \T', \P', N]$ will yield
equivalent quantum evolution equations of the same form as
\eqref{paramscheq}--\eqref{paramH} with the substitutions $\T
\rightarrow \T'$ and $\P\rightarrow \P'$. This suggests the set of
parametric quantum evolution equations is somehow more flexible in
choosing a `time' variable than that of the \sch{} equation
\eqref{scheq} that assumes the existence of a preferred choice of
time. A further advantage of the present approach is the
implication that the action \eqref{scheq3} for a wider class of
the kinematic actions with a general $\HQK = \HQK(\T,\P)$ in place
of \eqref{HKP} may play a fundamental role in the quantization of
a parametric theory where no preferred (Newtonian) time can be
identified. In this case the parametric quantum evolution system
\eqref{paramscheq}--\eqref{paramH} becomes nonlinear and cannot be
reduced to the \sch{} equation \eqref{scheq}. The following
sections will provide illustrative examples for this
generalization based on {minisuperspace} models.

\section{Friedmann universe with a quantized massless scalar field}
\label{frw}

The theoretical framework for the quantization scheme developed in
the previous section will now be employed in the quantization of a
simple Friedmann universe filled with a massless scalar field. The
nonlinear  quantization of a similar model with a massive scalar
field was addressed in \cite{wang_2003} in terms of a nonlinear
integro-partial differential system derived from
\eqref{paramscheq}--\eqref{paramH} (for $\alpha=0$), which was
reduced to a system of nonlinear ordinary differential equations
of infinite dimensions. The latter was further truncated for
numerical simulation. Not surprisingly, the reduced system can be
shown to be non-dissipative since there exists an underlying
variational principle as developed
in the previous section. 

The present Friedmann model with a massless scalar field can be
investigated analytically.
The classical Lagrangian for this model may be
found from \cite{Blyth_Isham_1975} and \cite{wang_2003}
by dropping the mass term to obtain
\begin{equation}\label{LFRW}
L(\phi, \dot\phi, R, \dot{R}, N)
=
-\frac{6 R }{N}\dot{R}^2 + \frac{6 R^3}{N}\dot{\phi}^2  + 6 N K R .
\end{equation}
Here  $\phi=\phi(t)$ is the scalar field $\phi$ which has been
conveniently re-scaled. The positive variables $R=R(t)$ and
$N=N(t)$ are the scale factor and lapse function respectively
appearing in the Robertson-Walker metric
\begin{equation}\label{RW}
\d s^2 = -N^2\d t^2 + R^2 \d\ell^2
\end{equation}
with $\d\ell^2$ denoting the squared line element on the
homogeneous and isotropic 3-space, which may be closed ($K=1$),
flat ($K=0$) or `open' ($K=-1$). The momenta conjugate to $R$ and
$\phi$ are
\begin{align}\label{}
\P &:= \pd{L}{\dot{R}} =  -\frac{12 R }{N}\dot{R} \\
p &:= \pd{L}{\dot{\phi}} = \frac{12 R^3}{N}\dot{\phi} .
\end{align}
It follows that the
corresponding Hamiltonian $H(\phi, p, R, \P, N) := p \,\dot{\phi} + \P \dot{R} - L$
takes the form $H = N\H$ where
\begin{equation}\label{HHFRW}
\H(\phi, p, R, \P)  = - \frac{\P^2}{24R} +  \frac{p^2}{24 R^3}  - 6 K R .
\end{equation}
The canonical equations of classical motion can be derived with
arbitrary $N(t)$. The gauge condition $N = 12 R^3$ is chosen so
that these equations take a simpler form as follows
\begin{equation}\label{class_phi}
\dot{\phi} = {p}
\end{equation}
\begin{equation}\label{class_p}
\dot{p}=0
\end{equation}
\begin{equation}\label{class_R}
\dot{R} = -R^2 \Pi
\end{equation}
\begin{equation}\label{}
\dot{\Pi} =
-\frac{R \Pi^2}{2}
+\frac{3 p^2}{2 R}
 + 72 K R^3
\end{equation}
subject to the (Hamiltonian) constraint
\begin{equation}\label{}
- \frac{\P^2}{24R} +  \frac{p^2}{24 R^3}  - 6 K R = 0.
\end{equation}
Clearly,
from \eqref{class_p} the momentum $p$ is a constant of motion,
with which \eqref{class_phi} is readily integrated.
With a suitable choice of the origin of the coordinate time $t$,
one obtains
\begin{equation}\label{R2}
R^2 = {\frac {\abs{p} \,{e^{2 \abs{p} t}}}{6(1+K{e^{4 \abs{p} t}})}}
\end{equation}
for $-\infty < t < \infty$ with $K = 0, 1$ and  $-\infty < t < 0$
with $K = -1$. In terms of this expression \eqref{class_R} can be
easily solved for $\Pi$.

The quantization of this model requires a degree of freedom to be
quantized. We choose this to be the scalar field $\phi$, thereby
regarding the scale factor as the `geometric time'. As per
discussions in section~\ref{parasch} we split $\H$ by
\begin{equation}\label{}
\H = \HK + \HD
\end{equation}
where
\begin{equation}\label{}
\HK = - \frac{\P^2}{24R}
\end{equation}
\begin{equation}\label{}
\HD = \frac{p^2}{24 R^3}  - 6 K R .
\end{equation}
To proceed
the operator
\begin{equation}\label{BKop}
\hat{h} =  -\frac{1}{24 R^3}\pd{^2}{\phi^2} - 6 K R
\end{equation}
is constructed by substituting
$p \rightarrow \hat{p} := - \im\pd{}{\phi}$ into $\HD$. This operator
is Hermitian with respect to the inner product
\begin{equation}
\label{FangPsi}
\ang{\Psi_1}{\Psi_2} := \int_{-\infty}^{\infty}  \!\Psi_1^* \Psi_2 \,\d\phi
\end{equation}
of any two wavefunctions $\Psi_1(\phi,t)$ and $\Psi_2(\phi,t)$.

In the gauge $N = 12 R^3$, the following nonlinear evolution
equations for the normalized wavefunction $\Psi(\phi,t)$ and
kinematic variables $R(t), \P(t)$ then arise from
\eqref{paramscheq}--\eqref{paramH}:
\begin{equation}\label{sch}
\im \pd{\Psi}{t}  =   -\frac{1}{2}\pd{^2\Psi}{\phi^2}
\end{equation}
\begin{equation}\label{dR1}
\dot{R} = -R^2 \Pi
\end{equation}
\begin{equation}\label{}
\dot{\Pi} = -\frac{R \Pi^2}{2} +\frac{3
\ave{\hat{p}^2}
}{2 R} + 72 K R^3
\end{equation}
subject to the constraint
\begin{equation}\label{H0}
 -\frac{\P^2}{24R} +\frac{\ave{\hat{p}^2}}{24 R^3} - 6 K R= 0
\end{equation}
where
\begin{equation}\label{}
\ave{\hat{p}^2}
= -\int_{-\infty}^{\infty} \Psi^* \pd{^2}{\phi^2} \,\Psi \,\d \phi
\end{equation}
and the phase condition $\alpha= - 6 K R$ has been used for
simplicity. {These equations are similar in structure
to those obtained by Kim in analyzing the Friedmann universe filled
with a massless scalar field in the context of semiclassical
gravity \cite{kim}.}

The wave equation \eqref{sch} in fact decouples from the others
and takes the same form as the 1-dimensional \sch{} equation for a
free non-relativistic particle. Hence $\ave{\hat{p}^2}$ is
preserved under evolution. It follows that the solution for $R$
can be generated from the classical solution \eqref{R2} by
replacing $\abs{p}$ with $\sqrt{\ave{\hat{p}^2}}$. Like the
classical case, there is no singularity avoidance and $R$ can
become arbitrarily close to zero for all possibilities of $K$.
Unlike the classical case, however, it is possible to envisage a
wave packet with zero mean scalar field and momentum
($\av{\phi}=\av{\hat{p}}=0$) but nonzero deviations ($\av{\phi^2}
> 0$ and $\av{\hat{p}^2} > 0$). In this case, the evolution of the
Friedmann universe can be thought of as being `purely
quantum-driven'.

\section{Nonlinear quantum cosmology of the Bianchi type}
\label{Bianchi}

The foregoing Friedmann universe with a scalar field
was quantized while maintaining the classical nature of
the scale factor which carries the gravitational
degree of freedom of the universe. In this section,
two vacuum cosmological models, of Bianchi types I and IX, will be examined.
Both belong to a wider class of
homogeneous but anisotropic spacetimes \cite{Ryan_1975}.
In Misner's seminal paper \cite{misner_1969},
the classical dynamics and quantization
of Bianchi I and IX universes in the context of
ADM quantization has been studied in detail.
Below we will give derivations of the
quantum evolution equations for
these two cosmological models and
comment on the consequences. The emphasis will be placed on
the nonlinearity inherent in the derived
quantum systems.

Just as in the Friedmann universe, the geometrical description
of the Bianchi  I and IX models involves the
lapse function $N(t)$ and
scale factor
$R(t)$. Two additional functions
$\beta_{+}(t)$ and
$\beta_{-}(t)$ are used, to allow for the dynamical
anisotropy of the spacial hypersurface at any coordinate time $t$.
Following Misner's notation we consider a spacetime metric of the form
\begin{equation}
\d s^2=-N^2 \d t^2 + R^2(e^{2 \beta})_{ij} \,\sigma_i
 \sigma_j
\end{equation}
for $i,j = 1,2,3$, where $\sigma_i$ denote
some basis 1-forms of the spatial hypersurface and $\beta$ is a
traceless matrix  with elements given by
\begin{equation}
\beta_{ij}={\rm diag}(\beta_{+} + \sqrt{3}\beta_{-},\
\beta_{+}- \sqrt{3}\beta_{-} ,\ -2\beta_{+}) .
\end{equation}
The constant time spacial hypersurface for the Bianchi  I universe
is open and has an Abelian homogeneity group. Thus
$\sigma_i$ are chosen so that
$\d \sigma_i = 0$ and can
be simply expressed as $\sigma_1 = \d x,\;\sigma_2 = \d
y,\;\sigma_3 = \d z$, for $-\infty < x, y, z < \infty$.
The homogeneity group of the closed spacial
hypersurface for the Bianchi IX  universe is that of ${\rm S}^3$.
Therefore the basis 1-forms  $\sigma_i$ are chosen to satisfy the
structure equations
\begin{equation}\label{}
\d \sigma_i= \frac{1}{2}~\epsilon_{ijk}~\sigma_j \wedge \sigma_k
\end{equation}
and can be
expressed explicitly   as
\begin{equation}
\sigma_1 = \sin{\psi}\d \vartheta- \cos{\psi} \sin{\vartheta}\d \varphi,\;
\sigma_2 = \cos{\psi}\d \vartheta+ \sin{\psi} \sin{\vartheta} \d \varphi,\;
\sigma_3 = -\d \psi - \cos{\vartheta}\d \varphi
\end{equation}
for  $0 \le {\vartheta} \le \pi$,  $0 \le \varphi < 2\pi$ and  $0 \le \psi < 4\pi$.

The classical dynamics of the models is generated by the Dirac-ADM
action subject to the homogeneity condition for all metric
functions. Up to an overall factor, the resulting Lagrangian takes
the form
\begin{equation}\label{LB}
L(\beta_\pm, \dot{\beta}_\pm, R, \dot{R}, N)
=
-\frac{6 R }{N}\dot{R}^2 + \frac{6 R^3}{N}
\left(\dot{\beta}_{+}^2
+\dot{\beta}_{-}^2\right)
+  N  R^3 \R
\end{equation}
where $\R$ denotes the scalar curvature on the spatial hypersurface. It can be conveniently expressed as
\begin{equation}\label{scr}
\R = \frac{3}{2 R^2}(1-V)
\end{equation}
where
\begin{equation}\label{V1}
V = 1
\end{equation}
for the Bianchi I model (such that $\R=0$) and
\begin{eqnarray}\label{V9}
V &=& V(\beta_{+},\beta_{-})= \nonumber \\
&&
\hspace{-8mm}
1
+\frac13\,e^{-8\beta_{+}}
+\frac13\,e^{4(\beta_{+}+\sqrt{3}\beta_{-})}
+\frac13\,e^{4(\beta_{+}-\sqrt{3}\beta_{-})}
-\frac23\,e^{4\beta_{+}}
-\frac23\,e^{-2(\beta_{+}+ \sqrt{3}\beta_{-})}
-\frac23\,e^{-2(\beta_{+}-\sqrt{3}\beta_{-})}
\end{eqnarray}
for the Bianchi IX model.
By introducing the conjugate momenta
\begin{eqnarray}
  \Pi &:=& \frac{\partial{L}}{\partial{\dot{R}}}= \frac{-12R}{N}\dot{R} \\
  p_{\pm} &:=& \frac{\partial{L}}{\partial
{\dot{\beta}_{\pm}}}= \frac{12R^3}{N}\dot{\beta}_{\pm}
\end{eqnarray}
we can construct the Hamiltonian
$H({\beta}_{\pm}, p_{\pm}, R, \P, N):=\Pi\dot{R}+p_+\dot{\beta}_+  + p_-\dot{\beta}_-  -L = N\H$
where
\begin{equation}\label{HHB}
\H({\beta}_{\pm}, p_{\pm}, R, \P)=
- \frac{\Pi^2}{24 R}+\frac{1}{24 R^3}\left({p^2_+ +p^2_-}\right)  + \frac{3R}{2} \,(V-1) .
\end{equation}
From \eqref{HHB} the canonical equations of motion can be derived and take a simpler form
using the choice $N = 12 R^3$ as in the Friedmann case. In this gauge,
denoting $\partial_\pm := \partial_{\beta_\pm}$,
one obtains:
\begin{equation}
\dot{\beta}_{\pm}= p_{\pm}
\end{equation}
\begin{equation}
\dot{p}_\pm = -18R^4
\partial_\pm{V}
\end{equation}
\begin{equation} \label{BP}
\dot{R}= -\Pi R^2
\end{equation}
\begin{equation} \dot{\Pi}=\frac{\Pi^2
R}{2}+\frac{3}{2{R}}\left({p^2_+ +p^2_-}\right)-18R^3(V-1)
\end{equation}
and the Hamiltonian constraint
\begin{equation}
- \frac{1}{2}\Pi^2 R^2+\frac{1}{2}\left(p_{+}^2 + p_{-}^2\right)  + 18 R^4 (V-1) = 0 .
\end{equation}
The analogy
between the Bianchi and Friedmann models
should be apparent. Specifically by
comparing the Lagrangian in \eqref{LB} with that in \eqref{LFRW} we see that
the kinetic terms in both cases share the same structure with the functions
$\beta_\pm$ resembling the scalar field $\phi$.

For the Bianchi  I model with $V=1$,
both $p_{+}$ and $p_{-}$ are constants of motion. By introducing
\begin{equation}\label{}
p := \sqrt{p_{+}^2 + p_{-}^2}
\end{equation}
the solution for $R$ may be generated from the scale factor
in the Friedmann model with zero curvature, i.e. $K=0$, given in
\eqref{R2}. This yields
\begin{equation}\label{BR2}
R^2 = {\frac {{p}}{6}}  \,{e^{2 {p}\, t}}
\end{equation}
for $-\infty < t < \infty$, from which $\Pi$ can be evaluated using \eqref{BP}.
The scale factor \eqref{BR2}
together with $N = 12 R^3$ and arbitrary constants $p_\pm$
is
equivalent to the Kasner solution  for the Bianchi  I universe  \cite{kasner_1921}.

The  quantization of the Bianchi models can be performed once
the dynamical variables to be quantized are decided. Guided by the analogous
roles played by $\beta_\pm$ and $\phi$, we choose to turn  $\beta_\pm$
into quantum variables and introduce their wavefunctions $\Psi(\beta_\pm,t)$.
The inner product of two such  wavefunctions $\Psi_1, \Psi_2$ can be defined
to be
\begin{equation}
\label{BangPsi}
\ang{\Psi_1}{\Psi_2} := \int_{-\infty}^{\infty}  \!\int_{-\infty}^{\infty}  \!
\Psi_1^* \Psi_2 \,\d\beta_+\d\beta_- .
\end{equation}
We are concerned only with normalized wavefunctions $\Psi$
satisfying $\ang{\Psi}{\Psi} = 1$ so that $\av{\hat{o}} = \ang{\Psi}{\hat{o}\,\Psi}$
for  any operator $\hat{o}$.

As before, the scale factor $R$ will act as the classical time
that evolves with respect to the coordinate time $t$. The coupling
between these quantum and classical variables follow from
\eqref{paramscheq}--\eqref{paramH} applied in the current setting.
In accordance with procedures leading to \eqref{sch}--\eqref{H0},
we perform the `kinematic-dynamic' split
\begin{equation}\label{}
\H = \HK + \HD
\end{equation}
where
\begin{equation}\label{}
\HK = - \frac{\P^2}{24R}
\end{equation}
\begin{equation}\label{}
\HD = \frac{1}{24 R^3}\left({p^2_+ +p^2_-}\right)  + \frac{3R}{2} \,(V-1)
\end{equation}
and construct the operator
\begin{equation}\label{Kop}
\hat{h} =  -\frac{1}{2}\left(\partial_+^2+\partial_-^2\right){\Psi}
 + 18 R^4 (V-1)
\end{equation}
from $\HD$ with the
substitutions
$p_\pm \rightarrow \hat{p}_\pm := - \im\partial_\pm$.

By using the conditions $N = 12 R^3$ and $\alpha=0$
we arrive at
the following
nonlinear evolution equations
\begin{equation}\label{Bsch}
\im \partial_t{\Psi}  =
-\frac{1}{2}\left(\partial_+^2+\partial_-^2\right){\Psi}
 + 18 R^4 (V-1) {\Psi}
\end{equation}
\begin{equation}\label{BdR1}
\dot{R}= -\Pi R^2
\end{equation}
\begin{equation} \dot{\Pi}=\frac{\Pi^2
R}{2}+\frac{3}{2{R}}\ave{\hat{p}^2_+ +\hat{p}^2_-}-18R^3(\ave{V}-1)
\end{equation}
subject to the constraint
\begin{equation}\label{BH0}
- \frac{1}{2}\Pi^2 R^2+\frac{1}{2}\ave{\hat{p}^2_+ +\hat{p}^2_-}  + 18 R^4 (\ave{V}-1) = 0
\end{equation}
for $\Psi(\beta_\pm,t)$, $R(t)$ and $\P(t)$.

For the quantized Bianchi  I model ($V=1$), the decoupled wave
equation \eqref{Bsch} has the same structure as the \sch{}
equation for a free non-relativistic particle in 2-dimensions as
expected. Hence $\av{\hat{p}^2_+ +\hat{p}^2_-}$ is preserved under
evolution and the solution for $R$ can be generated from the
classical solution \eqref{BR2} by replacing $p$ with
$\sqrt{\av{\hat{p}^2_+ +\hat{p}^2_-}}$. Clearly as $t\rightarrow
-\infty$ the scale factor tends to zero, i.e. $R \rightarrow 0$,
where the spacetime becomes singular. Using the above properties,
one may demonstrate explicitly how this universe may be driven by
quantized geometry described by a wave packet with zero means of
the dynamical gravitational fields and their momenta
($\av{\beta_\pm}=\av{\hat{p}_\pm}=0$)
but with nonzero deviations
($\av{\beta_+^2 +\beta_-^2}>0$ and $\av{\hat{p}^2_+
+\hat{p}^2_-}>0$).

For the quantized Bianchi  IX model where $V$ is given in
\eqref{V9}, equations \eqref{Bsch}--\eqref{BH0} constitute a
system of fully coupled {\em nonlinear} integro-partial
differential equations. In the absence of explicit solutions it is
not possible to analyze the dynamical behaviour of this model in
detail. Nonetheless, it is worth noting two generic properties
that one may conclude from this system. First, it can be shown
that $V$ is nonnegative with its minimum $V=0$ taking place at
$\beta_\pm=0$. (In fact $V \approx 8\beta_+^2+8\beta_-^2$ for
$\abs{\beta_\pm} \ll 1$. For further details of $V$,
see~\cite{misner_1969}.) It follows that the term $V-1$ in $\H$
can become negative which could be problematic for ADM
quantization due to a square root taking procedure. The proposed
quantization scheme is free from this problem, as  exemplified in
\eqref{Bsch}--\eqref{BH0} where no square roots are involved.
Secondly, although the scale factor is treated as a classical
variable it is not used as an evolution parameter. As such, the
quantum evolution will not come to a halt at the `maximum
hypersurface' where $\dot{R}=0$ classically. By comparison, the
pathology associated with time integration across such a
hypersurface could be suffered by both ADM quantization and Dirac
quantization using \kuchar's embedding variables
\cite{kuchar_1992}.

\section{Conclusions and discussions}
\label{fin}

By virtue of its action principle, we have reformulated the \sch{}
equation in non-relativistic quantum mechanics into a parametric
form. In this formalism the Newtonian time acts as a kinematic
variable that evolves with respect to a general time coordinate
while coupling to a quantum wavefunction in a semiclassical
fashion. A gratifying feature of this description is that the
parametrized action decomposes into a kinematic part and dynamic
part. The former part of the action depends only on classical
variables and takes a canonical form. Therefore, by preserving
this form, one may perform a canonical transformation that
redefines the kinematic time and its conjugate momentum. This
suggests a natural extension of the proposed quantization method
by accommodating a general expression of the kinematic time and
its conjugate momentum that enter into the kinematic action. A
quantum description obtained from this generalization is no longer
reducible to the standard \sch{} equation. Instead, it leads to a
theory for nonlinear quantum evolution previously proposed based
on heuristic arguments \cite{wang_2003}. In spite of the breakdown
of the principle of superposition in time evolution, the resulting
theory belongs to a class of nonlinear quantum theories admitting
both probabilistic interpretation and a Hilbert space with
Euclidean norm \cite{Mielnik_1974, Weinberg_1989a}.

These properties have indeed been demonstrated in the quantization
of three types of cosmological models in this paper. The first
model we considered was the Friedmann universe with a massless
scalar field. The model consists of both gravitational and  matter
degrees of freedom, carried by the scale factor and scalar field
respectively. Although classically these two variables take the
same status as dynamical variables, given the physical origin of
this model, we believe that it is matter that should be quantized
thereby making the scale factor the intrinsic geometric time. In
doing so we have not treated the scale factor merely as an
evolution parameter as in certain alternative quantization
methods. One obvious benefit of allowing the intrinsic geometric
time to evolve as a classical variable is so that the theoretical
description is not confined to scenarios with monotonically
ascending geometric time. This avoids the diverse choice of
time variables for different curvature parameters if the
same cosmological model was quantized using the ADM method
\cite{Blyth_Isham_1975}. It might be perceived at this point that
our approach only applies to the quantization of matter in
classical spacetime. To explicitly demonstrate it is not the
case, we then considered the Bianchi I and IX cosmological models
describing the evolution of an `empty' universe. No matter fields
are present there and one must decide which gravitational degrees
of freedom should be quantized. In section~\ref{Bianchi} the two
metric functions representing the anisotropy of the universe have
been chosen for this purpose. Arguably this is the most
appropriate choice due to the physical interpretation of these two
functions as the gravitational `waves' in the cosmological model.
These considerations suggest that in extending the proposed
quantization to quantum gravity, a primary concern is to determine
which  two gravitational field components are to be quantized so
that  the remaining four fields will act as classical kinematic
variables. Unlike \kuchar's {embedding} variables, however, our approach does
not require that the momenta of the kinematic variables to be resolved in
the Hamiltonian and momentum constraints. In contrast, it may be
speculated that if gravity were to be quantized based on the
proposed quantization then the corresponding  action integral
should contain a kinematic part of the form
\begin{equation}\label{}
S^{\rm K}[ X^A, P_A, N^A]
=
\int\!\d t \!\int\!\d ^3 x \left\{P_A \dot{X}^A -
{N^A \HK_A({X}^B,P_B)}
\right\}
\end{equation}
with $A,B = 0,1,2,3$  where $N^A$ are Lagrangian multipliers
proportional to the lapse and shift functions, $X^A$ and $P_A$
denote the gravitational fields and their momenta to act as
classical kinematic variables and {$\HK_A({X}^B,P_B)$} are some
expressions, {\em not necessarily linear in} $P_A$, to be
determined. {If such a split can be found uniquely in the full
theory, then the classification of the geometrodynamical variables
into kinematic and dynamic sets will follow naturally. In
contrast, the classifications of variables in the truncated
cosmological models in sections~\ref{frw} and \ref{Bianchi} have
been partly guided by physical intuition.}

A main thrust for this work has been the preservation of unitarity
in quantizing physical systems with generic nonlinear kinematical
structure. Of course, there exist other model-dependent strategies
for unitary evolution within the linear quantum formulation, e.g.
\cite{dust}. In a broader context, even the principle of unitarity
can be debated  by relaxing the probabilistic interpretation. Some
have {suggested} that it is at best an approximate concept
\cite{kim,kiefer}. However, a more conclusive argument on the
probabilistic interpretation of quantum gravity does not appear to
be available at present. The non-unitary Wheeler-DeWitt equation
{is} thought to be capable of describing a tunnelling
\cite{tunnel} or topology-changing  \cite{kim1999} wavefunction in
certain models of the early universe.
To what extent these theoretical scenarios can be accommodated or
precluded by our approach would form a subject for future investigation.

Although the proposed framework is somewhat reminiscent of
semiclassical gravity original developed by M{\o}ller and
Rosenfeld \cite{semicls} as an exact theory, where {\em all}
gravitational degrees of freedom are classical, it is worth
stressing that we seek to quantize {\em two} gravitational field
components. On the other hand, the remaining kinematic variables
in our approach are classical by construction and therefore do not
arise from a decoherence process required by Kiefer's approach to
semiclassical gravity as an approximation to quantum gravity
\cite{kiefer}.

Before further potential implications of this work on quantum
gravity can be contemplated seriously, several pressing issues
ought to be addressed first, regarding the extension of the
current quantum mechanical approach to a field-theoretical
description. In this respect, it would be profitable to investigate the
quantum dynamics of systems such as  scalar fields in curved
spacetime and
gravitational wave spacetimes \cite{gw}.  Progress
made along these lines will pave the way for the analysis of a
number of midisuperspace models where Dirac quantization using
\kuchar{}'s method has not been successfully applied. Results from
research described above will be reported elsewhere.


\section*{Acknowledgments}

We are most grateful to A Balanov, D A Burton, C J Isham, N
Janson, J Louko, R Maartens,  M A H MacCallum and R W Tucker for
fruitful conversations and helpful comments on aspects in general
relativity, quantum gravity, cosmology and nonlinear dynamical
systems. Thanks are also due to the referees who {drew} our
attention to works in \cite{HK, dust, bar, kim, kiefer, tunnel,
kim1999}. The  research has been carried out under partial
financial support from the EPSRC.

\end{document}